\documentclass[preprint2]{aastex}
\usepackage{natbib,makeidx,psfig}
\usepackage{lscape}

\newcommand{\Ha}{\mbox{H$\alpha$}}
\newcommand{\Hb}{\mbox{H$\beta$}}
\newcommand{\etal}{\mbox{et al.}}
\newcommand{\ie}{\mbox{i.e.}}

\newcommand{\vi}{\mbox{$V\!-\!I$}}

\received{2001 November 7}
\begin{document}

\slugcomment{Submitted for publication in {\it The Astrophysical
Journal}}

\title{Spectrophotometric Observations of Blue Compact Dwarf Galaxies:  
Mrk~370}

\author{Luz M. Cair\'os}
\email{luzma@uni-sw.gwdg.de}
\affil{Universit{\" a}ts-Sternwarte G{\" o}ttingen, 
Geismarlandstr. 11, 37083, G{\" o}ttingen, Germany. 
Instituto de Astrof\'\i sica de Canarias , E-38200  La Laguna, Tenerife,
Canary Islands, Spain. 
Departamento de Astronom\'\i a, Universidad de Chile, Casilla 36-D, Santiago, 
Chile}

\author{Nicola Caon}
\email{ncaon@ll.iac.es}
\affil{Instituto de Astrof\'\i sica de Canarias, E-38200  La Laguna, Tenerife,
Canary Islands, Spain}

\author{Bego\~na Garc\'\i a Lorenzo}
\email{bgarcia@ing.iac.es}
\affil{Isaac Newton Group of Telescopes (ING), E-38780 Santa Cruz de La Palma,
La Palma, Canary Islands, Spain}

\author{Jos\' e M. V\'\i lchez}
\email{jvm@iaa.es}
\affil{Instituto de Astrof\' \i sica de Andaluc\'\i a, CSIC, Apdo. 3004,
18080 Granada, Spain}

\author{Casiana Mu\~noz-Tu\~n\'on}
\email{cmt@ll.iac.es}
\affil{Instituto de Astrof\'\i sica de Canarias, E-38200  La Laguna, Tenerife,
Canary Islands, Spain}

\accepted{}
\shortauthors{Cair\'os et al.}
\shorttitle{Spectrophotometric observations of Mrk~370}

\begin{abstract} 

We present results from a detailed spectrophotometric analysis of the blue
compact dwarf galaxy (BCD) Mrk~370, based on deep $UBVRI$ broad-band and 
\Ha\ narrow-band observations, and long-slit and two-dimensional spectroscopy 
of its brightest knots. 

The spectroscopic data are used to derive the internal extinction, and to
compute metallicities, electronic density and temperature in the knots. By
subtracting the contribution of the underlying older stellar population,
modeled by an exponential function, removing the contribution from emission 
lines, and correcting for extinction, we can measure the true colors of the 
young star-forming knots.  We show that the colors obtained this way differ
significantly from those derived without the above corrections, and lead to
different estimates of the ages and star-forming history of the knots. Using
predictions of evolutionary synthesis models, we estimate the ages of both the
starburst regions and the underlying stellar component. We found that we can
reproduce the colors of all the knots with an instantaneous burst of star 
formation and the Salpeter initial mass function ({\sc imf}) with an upper 
mass limit of 100 M$_{\odot}$. The resulting ages range between 3 and 6
Myrs. The colors  of the low surface brightness component are consistent with
ages larger than 5 Gyr. The kinematic results suggest ordered motion around
the major axis of the galaxy.

\end{abstract}

\keywords{galaxies: dwarf galaxies - galaxies: starburst - galaxies: compact -
galaxies: kinematics and dynamics -  galaxies: evolution}

\section{Introduction}

Blue Compact Dwarf galaxies (BCD) are low-luminosity (M$_{B} \geq -18$ mag),
compact (starburst diameter $\leq 1$ kpc) objects, which have optical spectra
similar to those presented by \ion{H}{2} regions in spiral galaxies
\citep{sargent70,Thuan81}. They are metal deficient galaxies ---~the
metallicities of their ionized gas ranging between Z$_{\odot}/50$ and
Z$_{\odot}/2$~---  that form stars at high rates, able to exhaust their 
available gas content in a time much smaller than the age of the Universe.
This fact implies that either these galaxies are young systems, experiencing 
their first burst of star formation, or that star formation occurs in short 
bursts separated by long quiescent periods \citep{sargent70}. 
Nowadays, the latter explanation is the more widely accepted, at least for 
the majority of BCDs \citep{thuan91}. 

In the recent years BCDs have attracted a great deal of interest, and have 
become key objects for understanding fundamental astrophysical problems.  Much
effort has been devoted to the study of these galaxies, driven by the following
fundamental questions: 1) what is the mechanism that triggers the starburst
activity and is there a unique one in all BCD subtypes?  2) what are the
properties of the underlying stellar population (age, metallicity, dynamical
status)? Answering the above questions would allow to form a general view of
dwarf galaxy formation and evolution, and in particular to test those
evolutionary  scenarios linking the different classes of dwarf galaxies,
namely, dwarf irregulars (dIs), dwarf ellipticals (dEs) and BCDs 
\citep{thuan85,davies88,papaderos96b,marlowe97,marlowe99}, or the different 
morphological classes of BCDs \citep{noeske00}. However, most of the work
carried out so far has focused on a statistical analysis of BCD samples by
means of surface photometry 
\citep{papaderos96a,telles95,doublier97,doublier99,oestlin98}, and has rather 
demonstrated the complexity of the topic than given firm answers.

Prompted by this lack of conclusive results, we started an extensive study of a
large sample of BCDs. The first step was the analysis of broadband 
observations in $B$, $V$, $R$ and $I$ for a sample of 28 BCDs. In 
\citet[hereafter Paper~I]{Cairos01a} we presented deep contour maps in the $B$ 
band, and surface brightness and color profiles, which were used to examine and
discuss their morphology and structure. In 
\citet[hereafter Paper~II]{Cairos01b} we enlarged the dataset by including  
$U$ and \Ha\ observations; we provided integrated photometry of the galaxies
and produced an atlas of detailed color and \Ha\ maps. The quality of the data 
allowed us to identify the different star-forming regions, and discriminate 
them from the regions where the star formation has already ceased.

One of our main conclusions was that optical broad-band photometry alone does
not allow to disentangle the distinct stellar components in BCDs and derive 
their evolutionary status. The measured broad-band colors are strongly 
affected by interstellar reddening and gaseous emission; besides, we need to 
separate the light coming from young stars from the emission of those
born in previous star-formation episodes ---~a methodological aspect often
overlooked.  Spectroscopic information is required in order to correct for
interstellar reddening and emission line contamination, and $U$-band and
\Ha\  measurements are fundamental, since they are much more sensitive to the 
star-formation properties.  Nevertheless, such kind of analysis is scarce in
the literature, mainly  because it requires a lot of observing time.  In
particular, not enough effort has been devoted so far to the characterization
of the stellar population hosting the actual burst, indeed a challenging task.

For these reasons, we decided to perform a more detailed study of a number of 
individual galaxies, based on a new strategy that combines broad- and
narrow-band photometry with long-slit and 2-D spectroscopy. The objects were
selected among those presented in Papers~I and II, so as to form
a sample including the different subtypes of objects classified as BCDs 
---~a not so well defined class which spans a wide range of luminosities, 
metallicities and morphologies \citep{kunth00, Cairos00, Cairos01b}.

This paper is the first in a series which present the results obtained from 
the use of this technique to analyze the selected galaxies. Here we describe 
the technique and its application to the galaxy  Mrk~370, a luminous BCD, with 
$M_{B} = -17.20$ (Paper~II), located at a distance of 12.9 Mpc 
\citep{Thuan81}\footnote{For consistency with Papers I and II, we use
here the distance from \citet{Thuan81}, which is based on the radial velocity 
with respect to the Local Group and $H_0 = 75$ km sec$^{-1}$ Mpc$^{-1}$; 
applying the correction for LG infall into Virgo would give $D = 10.9$ Mpc, 
and total luminosities, absolute magnitudes and sizes would change 
accordingly.}. 

We have chosen this galaxy in order to study a prototype of the group of BCDs 
populating the higher luminosity and metallicity range within the most common 
BCD morphological subclass, the iE type \citep[ hereafter LT86]{Loose86}. 
This type is characterized by an underlying, low brightness elliptical 
component which hosts numerous star-forming regions. 

In Mrk~370, the brighter knots of star-formation are located in the inner
part of the galaxy, from where smaller star-forming knots emerge and form a 
structure resembling spiral arms (see Papers~I and ~II).  The choice of Mrk~370
rests also on the fact that the comparison of the properties of its many
individual starburst knots provides valuable information on the mechanisms
that trigger the star formation, allowing to study how it propagates within
the galaxy, and is a possible test of the stochastic self-propagating star
formation mechanism \citep{gerola80}, which has been suggested to explain how
the star-formation process in dwarfs is maintained \citep{coziol96}.

In addition, Mrk~370 may be seen as a prototype of the BCDs classified as
multiple nucleus starburst galaxies. The presence of two central star-forming
knots led \citet{MazBor93} to label this galaxy as a "double nucleus" BCD,
prompting further suggestions of its merger origin. As we will show in this
work, the kinematic and spectrophotometric results obtained cast serious 
doubts on the merger scenario.

A complete collection of color and \Ha\ maps  of this galaxy can be found at
the following Web page:  \url{http://www.iac.es/proyect/GEFE/BCDs}. The basic
data of Mrk~370 are shown in Table~\ref{Tab:Galcoord}.

\section{Observations and Data Reduction}

\subsection{Long-slit spectra}

Long-slit spectroscopy of Mrk~370 was obtained in September 1998, with the
Isaac Newton Telescope (INT) at the Observatorio del Roque de Los Muchachos
(ORM, La Palma). The data were taken using the Intermediate Dispersion
Spectrograph (IDS), equipped with a CCD. A slit $3\farcm8$ long and
$1\farcs2$ wide was used. The position angle of the slit was set to
$125\arcdeg$, so as to pass through the two brightest central knots.
The spectral sampling was 1.85 \AA\ per pixel, the wavelength range 
$3600-7300$ \AA, and the spectral resolution $\sim 5.5$ \AA\ (FWHM); the 
spatial sampling was 0.39 arcsec per pixel. The seeing was about $1\farcs5$.
The log of the observation is shown in Table~\ref{Tab:Obslog}.

Data reduction was performed using IRAF standard tasks: after subtracting the 
bias and flat-fielding the spectra, they were calibrated in wavelength. The sky
spectrum was derived by averaging the signal in two windows 30--40 pixel wide 
outside the region where the object is still detectable, and subtracted out. 
The spectra were then corrected for atmospheric extinction and calibrated in 
flux by means of observations of spectrophotometric standards.

\subsection{Two-dimensional spectroscopy}

The 4.2m William Herschel Telescope (WHT, at the ORM, La Palma), equipped
with the fiber system INTEGRAL \citep{arribas98} and the WYFFOS spectrograph 
\citep{bingham}, was used to observe the central region of Mrk~370.

\begin{figure}[H]
\centerline{\psfig{figure=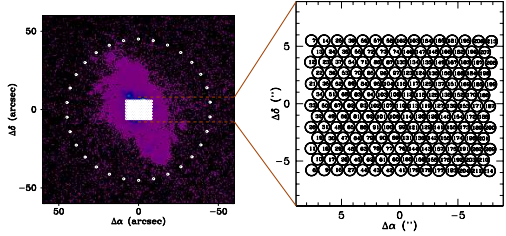,bbllx=49pt,bblly=00pt,bburx=1281pt,bbury=585pt,clip=,width=7.0truecm}}
\caption{({\it left}) Image of Mrk 370 through the \Ha\ filter, taken with 
the 2.2m telescope at CAHA. The spatial distribution of the fibers (SB2) on 
the focal plane has been overlaid. {\it (right)} Fibers in the central array. 
Numbers indicate the actual position of the fibers at the slit.}
\label{bundleSTD2}
\end{figure}

INTEGRAL currently has three fiber bundles with different spatial 
configurations on the focal plane. These three bundles can be interchanged 
during the observations depending on the scientific program or the seeing 
conditions. On the focal plane, the fibers of each bundle are arranged in two 
groups: one forms a rectangle which maps the object, the other forms an outer 
ring which samples the sky-background. 
Mrk~370 observations were done with the standard bundle 2 (hereafter SB2). 
It consists of 219 fibers, each $0\farcs9$ in diameter; the central rectangle 
is formed by 189 fibers covering an area of $16''\times12\farcs3$, while the 
other 30 fibers are evenly spaced on a ring $90''$ in diameter. 
Figure~\ref{bundleSTD2} shows the actual distribution of those fibers in the
focal plane. At the entrance of the WYFFOS spectrograph, the fibers are aligned
so as to form a pseudo-slit.

We observed Mrk~370 on August 22 1999, with a seeing of $\sim1\farcs3$.
The spectrograph was equipped with a grating of 600 groove/mm, and a Tek6 CCD
array of 1124x1124 24-micron pixels. The spectral range coverage was 
$4400-7150$ \AA,  with a spectral resolution of about 6 \AA; it includes
the \Hb, [\ion{O}{3}], \Ha, [\ion{N}{2}], and [\ion{S}{2}] emission lines (see
Appendix A). 

Data reduction was performed within the IRAF environment, and includes bias
and scattered light subtraction, fiber extraction, fiber throughput
correction, wavelength calibration, rejection of cosmic ray events, and 
correction from differential atmospheric refraction effects (for details on 
reduction of fiber data see e.g. \citealp{garcia99}, and references therein). 

In Appendix A we present the nuclear spectrum of Mrk 370 in the full wavelength
range observed with the fiber system.  We also show the individual spectra
corresponding to each of the 189 observed positions (fibers) in a selected 
spectral interval which includes the most important emission lines.

\subsection{H$\alpha$ imaging}

Narrow-band images centered on the \Ha\ line and on the adjacent continuum
were taken in 1997 December at the 2.2-m telescope of the German-Spanish
Astronomical Observatory on Calar Alto (Almer\'{\i}a, Spain).  The
instrumentation consisted of the {\em Calar Alto Faint Object Spectrograph},
CAFOS, and a $2048 \times 2048$ SiTe CCD chip. With a pixel size of 0.53 
arcsec, it provides a field 16 arcmin wide; however, due to the physical size 
of the available filters, only the central round area of about 11 arcmin is 
free from vignetting. The average seeing was 1.5 arcsec.

The image reduction was carried out using IRAF. Each image was corrected for 
bias, using an average bias frame, and was flattened by dividing by a mean 
twilight flatfield image. The frames were then registered (for each filter 
we took a set of dithered exposures) and combined to obtain the final frame, 
with cosmic ray events removed and bad pixels cleaned out. The average sky 
level was estimated by computing the mean value within several boxes 
surrounding the object, and subtracted out as a constant.
Flux calibration was done through the observation of spectrophotometric stars 
from the lists in \citet{oke90}. 

The complete log of the observations is shown in Table~\ref{Tab:Obslog}.

\subsection{Broadband Imaging }

UBVRI imaging of Mrk~370 was carried out in October 1998 at the 1-m Jacobus
Kapteyn Telescope, JKT (ORM, La Palma). We collected CCD images at the f/15
Cassegrain focus, using a Tek $1024\times1024$ pixel chip, yielding a pixel
size of 0.33 arcsec and a field of view of $5.6\times5.6$ arcmin. The
average seeing was $1\farcs5$. 

A log of the observations is given in Table~\ref{Tab:Obslog}. The image 
processing was performed using standard procedures available in IRAF.  
Flux calibration was done through the observation of photometric stars from 
\citet{landolt92} throughout the night.

\section{Spectroscopic Analysis} 

\subsection{Long-slit spectroscopy}

The position of the slit we used is shown in Figure~\ref{Fig:mkn370sli},
overplotted on a contour map of the continuum-subtracted \Ha\ image. 
(The meaning of the labels is explained in Section 4). The spatial run of the 
intensity of the brightest emission lines (\Ha, \Hb, [\ion{O}{3}]) has been 
used as a base for defining the regions from where the one-dimensional
spectra of each knot were extracted by integrating the signal within them.
Two distinct starburst regions have been identified, marked {\sc a} and 
{\sc b} in Figure~\ref{Fig:mkn370sli}. 

Figure~\ref{spec370} shows the summed spectra of the two knots, and the 
spectrum integrated over the whole galaxy (which we shall refer to as {\sc s}).
The spectrum of region {\sc a} shows a high continuum level. Absorption wings
are resolved in \Hb\ and H$\gamma$, H$\delta$ being visible in absorption
only. The Balmer discontinuity, together with these absorption features, can
be interpreted as the product of a substantial population of older stars,
with ages considerably larger than 10 million years. Region {\sc b} shows an
almost flat spectrum, characteristic of a dominant OB population, with no
evident absorption features. The integrated spectrum, {\sc s}, is similar to
that of region {\sc a}.

Fluxes and equivalent widths of the emission lines were measured using the
gaussian profile fitting option in the iraf task {\sc splot} (a direct
integration of the flux inside each line gave very similar results).
It is known that these measurements are an underestimate of the real flux
of the Balmer lines, because of the underlying absorption component.
In the present spectra we could not deblend the two components. To correct for
the underlying stellar absorption, some authors \citep{mccall85, skillman93}
adopt a constant equivalent width ($1.5-2$ \AA) for all the hydrogen 
absorption lines. However, the actual value of their equivalent width is 
uncertain and depends on the age of the star formation burst 
\citep{diaz88,cananzi93,olofsson95}. Thus we preferred to leave it as a 
free parameter, which was determined as follows.

We assume that the equivalent width, $W_{\rm abs}$, of the absorption line is
the same for all the Balmer lines, but can vary from knot to knot.
We then start with an initial guess for $W_{\rm abs}$, correct the measured 
fluxes, and determine the extinction coefficient at $\lambda=4861$ \AA, 
$C(\Hb)$, and its uncertainty, $\Delta C(\Hb)$, through a fit to the Balmer 
decrement, given by the equation:

\begin{equation}
\label{redde}
\frac{F(\lambda)}{F(H\beta)}=\frac{I(\lambda)}{I(H\beta)} \times 10^{C(H\beta)
\times f(\lambda)} 
\end{equation}

\noindent
where $\frac{F(\lambda)}{F(H\beta)}$ is the line flux corrected for absorption 
and normalized to \Hb;  $\frac{I(\lambda)}{I(H\beta)}$ is the theoretical
value for case B recombination, from \citet{brocklehurst71}, and $f(\lambda)$ 
is the reddening curve normalized to \Hb\ which we took from 
\citet{whitford58}.

We then search, for each knot, the value of $W_{\rm abs}$ that provides the
best match ---~that is, the lowest $\Delta C(\Hb)$~--- between the corrected 
and the theoretical line ratios.

Reddening-corrected intensity ratios and equivalent widths are quoted in 
Table~\ref{Tab:flux}. 

The [\ion{O}{3}] $\lambda\;4363$ line could not be detected in the spectra, so
the oxygen abundance for the different regions of Mrk~370 was derived using the
empirical 12+log O/H vs $R_{23}$ relation, first derived by \citet{pagel79}. We
have used the $R_{23}$ calibration according to \citet{mcgaugh91}. The ratio
[\ion{N}{2}]/[\ion{O}{2}] was used to break the degeneracy in the  calibration
in order to select either the upper or the lower branch  \citep{mcgaugh91}.
Since the values of [\ion{N}{2}]/[\ion{O}{2}] for the three spectra analyzed
are always above the 0.1 threshold ($\log \lambda 6584/\lambda 3727 = -0.59$,
$-0.81$, $-0.75$ for regions {\sc a}, {\sc b} and {\sc s}), the upper  branch
of the calibration has been used for the determination of the abundance. We
have derived the following values: $12+\log O/H = 8.7$, 8.6 and 8.7 for the
spectra of regions {\sc a}, {\sc b} and {\sc s} respectively, for which the
corresponding values for the logarithm of electron density are  2.07, 2.2  and
2.3.

The average value for the nitrogen to oxygen ratio is $\log N/O = -1.1$. 
The derived average oxygen abundance is approximately $Z_{\odot}$/2.
The abundances are equal for regions {\sc a} and {\sc b}, within the 
uncertainties.

\subsection{Two-dimensional spectroscopy}

Two-dimensional spectroscopy with optical fibers allows to collect
simultaneously the spectra of many different regions of an extended object, 
combining photometry and spectroscopy in the same dataset. 

From the individual measurements (line fluxes and ratios, continuum level,
radial velocity, etc.) obtained at each fiber, we can build 2-D maps by using
a two-dimensional interpolation technique, such as the Renka \& Cline
algorithm (implemented in the E01SAF and E01SBF routines in the NAG Fortran
Library).
This requires a precise knowledge of the exact location of each fiber, which
is determined by a metrology machine. In this way we built up images of
$95\times95$ pixels with a scale $\sim 0\farcs2$ pixel$^{-1}$. While the
spatial sampling (that is, the fiber diameter) of INTEGRAL (STD2) is 0.9
arcsec, the centroid of any peak in our maps can be measured with an accuracy
of around 0.2 arcsec (that is $~1/5$ of the fiber diameter; see for instance 
\citealp{Mediavilla98})

\subsubsection{Continuum and Line-Intensity Maps}

Because of non-photometric conditions during INTEGRAL observations, we could 
not flux-calibrate directly our spectra. The flux calibration was done
by using the long-slit data as a reference. We integrated the signal in our 
2-D spectra within two regions matching regions {\sc a} and {\sc b} in the 
long-slit spectra. Comparing the fluxes of different lines, we could compute 
the conversion factor with an accuracy of about 20\%. 
This is the error associated to the fluxes derived from INTEGRAL data using 
this calibration.

Figure \ref{continua}(a) shows a continuum map (within a spectral region where 
no emission lines are present) of Mrk~370 from INTEGRAL spectra.
We also present in Figure~\ref{continua}(b) a $V$ band image of the galaxy, in 
order to evidence the excellent agreement between the continuum map and the 
broad-band images.

The INTEGRAL data show that the position of the peak in the continuum maps 
does not depend on what spectral range was used to build them.
Therefore, we can safely define this point as the "optical nucleus" of the 
galaxy, and we set it as the origin of the coordinate system in all our 
spectral maps.

By fitting a single-gaussian to each emission line in our spectra, we have 
obtained emission line flux maps. Figure~\ref{continua} (c) shows the \Ha\ 
emission map. 
We also fitted an absorption component to Balmer lines in those spatial 
positions where the absorption wings were clear. For comparison, an \Ha\ 
narrow band filter image of Mrk~370 is displayed in Figure~\ref{continua}~(d).
The area covered by INTEGRAL includes four of the many star-forming knots 
present in the galaxy. The optical nucleus does not coincide with any of them.

Figure~\ref{continua} (e) is the map of the [\ion{O}{3}]/\Hb\ ratio, which is
commonly used as an indicator of the excitation. The morphology of the 
[\ion{O}{3}]/H$\beta$ map is similar to the emission lines intensity maps. 
The regions with the highest ionization coincide with the knot peaks, and the 
level of excitation is normal for star-forming regions. 
Figure~\ref{continua} (f) is the map of the extinction coefficient $C(\Hb)$ 
computed from the observed \Ha/\Hb\ ratio.

The peaks in the extinction coefficient map are close to, but not coincident 
with, the peaks of the \Ha\ knots. The average offset for the four knots in 
our observed area is $0.26\pm0.11$ arcsec. 
While this displacement is small ---~in fact, it is at our detection limit~---,
it may suggest a real spatial shift between the center of the HII regions and 
the {\it holes} in the C(\Ha) distribution.
A similar case has been reported before by \citet{maiz98} for the starburst
knots of the galaxy NGC 4214. They find the dust to be concentrated at the 
boundaries of the ionized regions. The explanation that they propose is
that stellar winds have contributed to diffuse the dust into the inter-cluster
medium.

\subsubsection{Kinematical Pattern}

The kinematic analysis is limited to the emission features, as absorption 
lines are too weak to be of some utility.

\noindent
{\em Individual Radial Velocities}

Radial velocities were obtained by fitting a single gaussian to each emission 
line present in our spectra. Although the spectral resolution of the Integral 
Field Spectroscopy is $\sim275$ km/s, the centroid of the emission features 
could be measured with an accuracy of $\pm10$ km/s, as indicated by the 
scatter among the radial velocities obtained from different lines. 
The final radial velocities of the ionized gas were obtained by averaging
the results derived from individual emission lines. We compared the results 
for the two different individual exposures of Mrk~370. The mean difference 
among the individual values was $0\pm12$ km/s. Therefore, by adding the error 
on the wavelength calibration ($\sim7$ km/s), we estimate that the final 
uncertainties on the velocity measurements are $\lesssim 20$ km/s.

\noindent
{\em The Mean Ionized Gas Velocity Field}

Figure~\ref{velocity_map} shows that the velocity field appears to be regular 
and similar to that of a rotating disk. The west region is receding, the
east part is approaching. In order to determine the location of the kinematic
center ({\sc k} hereafter), we computed the "differential" of the velocity
field, averaging the absolute differences between one point and its nearest
neighbors. This procedure emphasizes regions in which the velocity is changing 
rapidly. 
The position derived in this way is placed $\sim2\farcs4$ SE 
($-2\farcs36$, $0\farcs37$) of the optical nucleus. {\sc k} is located between
regions {\sc a} and {\sc b}, in an apparently  non-significant location in the
galaxy, and has a velocity value of 766$\pm20$ km/s, consistent with the
systemic velocity of the galaxy, 783$\pm16$ km/s, reported by \cite{Falco00}.

{\sc k} is nearly coinciding with the geometrical center of the external
isophotes, obtained from our broad band images.  This fact might suggest that
the star-forming knots in the central region of Mrk~370 could share the same
velocity field of the main body.

\noindent
{\em Velocity Field Parameters}

To quantify the relevant parameters of the ionized gas velocity field, we
adopted a simple kinematic model \citep{mihalas81}:
\begin{equation}
V_{\rm r}(R,\theta)=V_{\rm sys}+Z(R) \cos i + \Omega(R) \cos[\theta - \delta(R)]
\sin i
\end{equation}

where V$_{\rm sys}$ is the systemic velocity, $\Omega(R)$ is the strength of
the velocity field projection onto the galactic plane, and $\delta(R)$ is
the PA of the major kinematic axis. R and $\theta$ are the galactocentric polar
coordinates in the plane of the galaxy. 

We fit this model to the data, keeping fixed the position of the center at
{\sc k}, and assuming an initial estimate of the position angle of the 
kinematical major axis $\delta_{0} \sim 90^\circ$, and an inclination of the
galaxy $i \sim 40^\circ$ \citep{nordgren}.
The results do not depend on $\delta_{0}$, and are rather insensitive to
reasonable changes in {\it i}.
In Figure~\ref{modelo_mean} we present the dependence on galactocentric
distance of the parameters resulting from the fit. The systemic velocity we 
derived is 761 km/s, with an internal error of $\pm2$ km/s.

\section{Photometric Analysis}

\subsection{H$\alpha$ photometry of the starburst knots}

The left panel of Figure~\ref{Fig:mapas} shows a grey-scale \Ha\ image of 
Mrk~370. The galaxy reveals a complex structure consisting of several resolved 
knots located atop an extended low surface brightness (LSB) component having 
regular appearance (see the R band map in Figure~\ref{Fig:mapas}, right panel) 
and substantially redder colors (Paper~I and Paper~II). 

\begin{figure}[H]   
\centerline{\psfig{figure=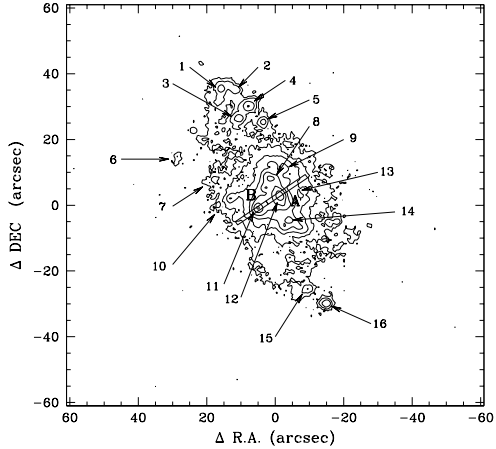,width=7.0truecm}}
\caption{Contour plot of the continuum subtracted \Ha\ image of Mrk~370.
The slit position is indicated, and the two subregions selected in the
long-slit spectrum are marked {\sc a} and {\sc b}. The individual star-forming 
regions are labeled. North is at the top and East is on the left.}
\label{Fig:mkn370sli}
\end{figure}

\begin{figure}[H]   
\centerline{\psfig{figure=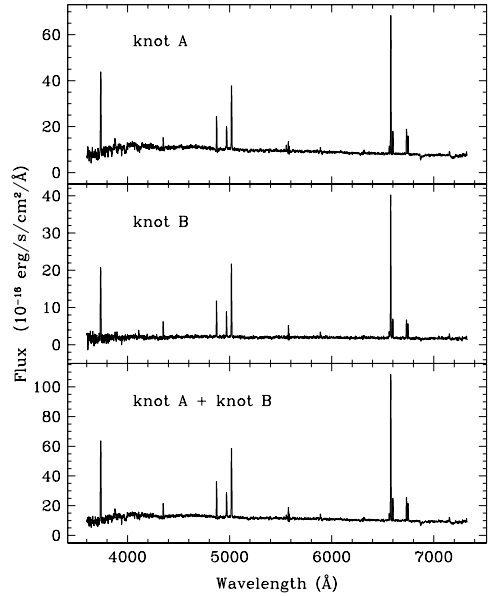,width=7.0truecm}}
\caption{Spectra of regions {\sc a} and {\sc b}, and the total integrated 
spectrum (region {\sc s}).}
\label{spec370}
\end{figure}

\begin{figure*}[H] 
\centerline{\psfig{figure=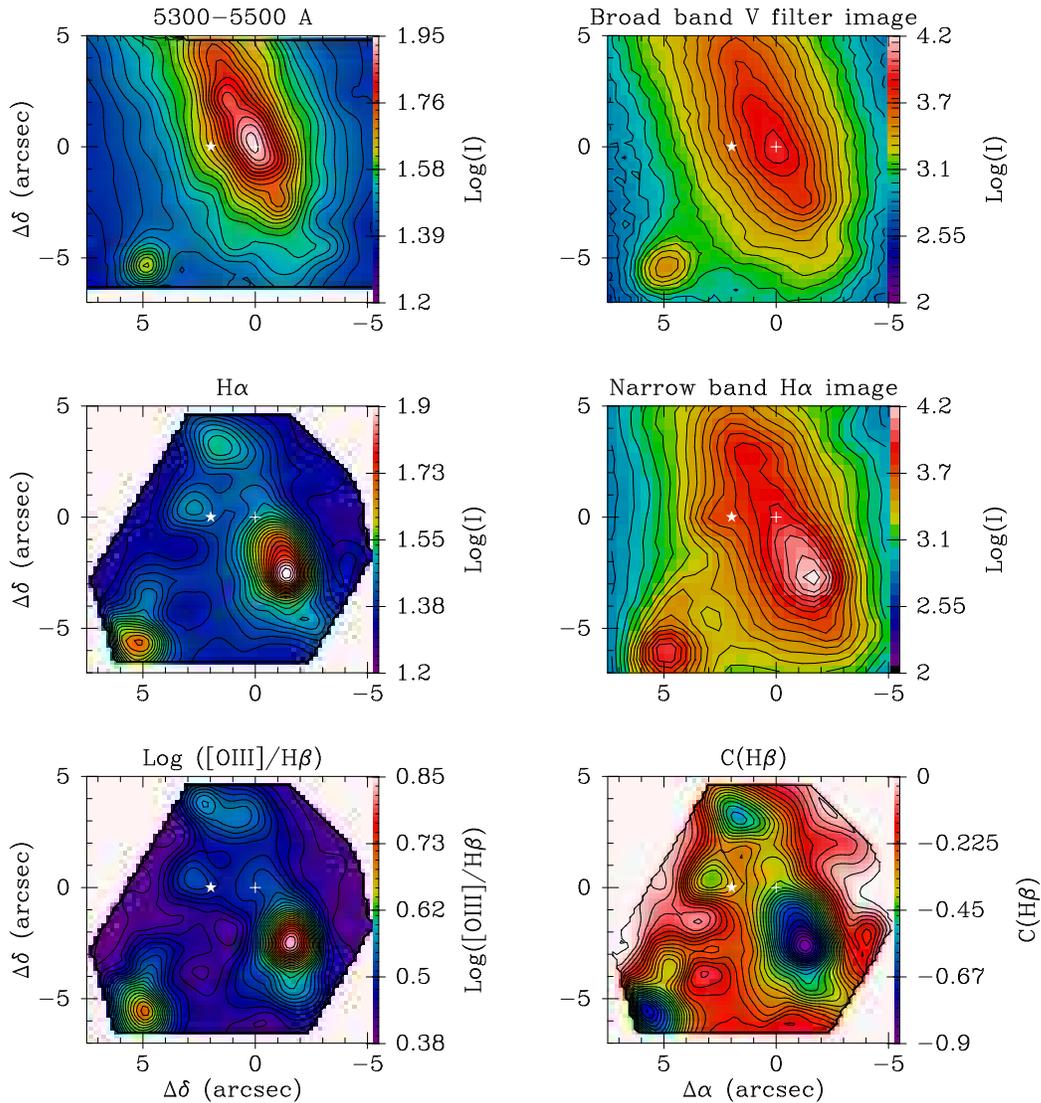,bbllx=0pt,bblly=100pt,bburx=572pt,bbury=696pt,clip=,angle=-90,width=16.0truecm}}    
\caption{[from left to right, top to bottom] 
(a) Mrk~370 continuum map obtained from the 2D spectroscopic data by
integrating the signal in the indicated spectral interval; 
(b) V filter image;
(c) Intensity map of the \Ha\ emission line obtained by fitting a single
gaussian to 2D spectroscopy data; (d) \Ha\ filter image from the 2.2m CAHA; 
(e) Two-dimensional distribution of [\ion{O}{3}]/\Hb\ ratio;
(f) Two-dimensional distribution of the extinction coefficient $C(\Hb)$.
A cross marks the optical nucleus in each map. The geometrical center of 
the outer isophotes is indicated by a star.}
\label{continua}
\end{figure*}

\begin{figure}[H]                         
\centerline{\psfig{figure=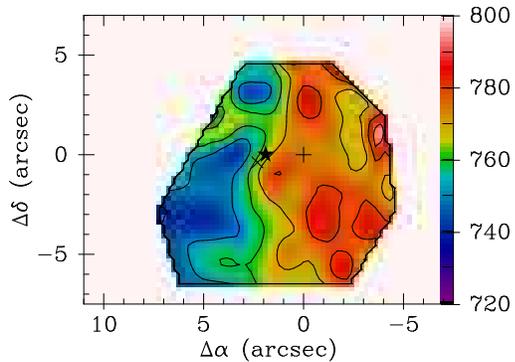,bbllx=40pt,bblly=49pt,bburx=572pt,bbury=696pt,clip=,angle=-90,width=7.0truecm}}
\caption{Velocity field of the ionized gas in the central region of Mrk~370, 
determined by averaging measurements from different emission lines. 
A black cross marks the optical nucleus. The kinematic center is indicated by 
an X. The center of the outer isophotes is indicated by a black star.}
\label{velocity_map}
\end{figure}

\begin{figure}[H]                 
\centerline{\psfig{figure=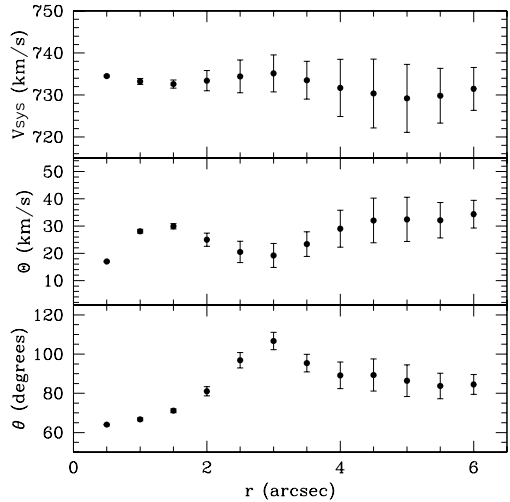,width=7.0truecm}}
\caption{Radial dependence of the systemic velocity V$_{\rm sys}$, amplitude 
of the velocity field projection on the galactic plane $\Omega$, and position 
angle of the major kinematical axis $\delta$.
Error bars only reflect the local {\it rms} in the measurements, and do not
reflect the actual uncertainties, which are likely affected by systematic 
errors.}
\label{modelo_mean}
\end{figure}

\begin{figure}[H]
\centerline{\psfig{figure=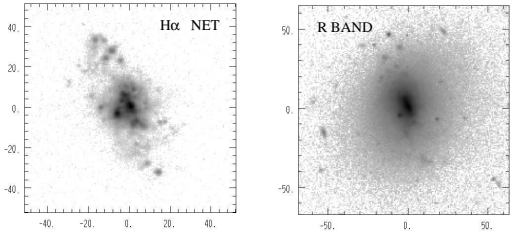,width=7.0truecm}}  
\caption{{\em left panel}: Continuum-subtracted \Ha\ grey-scale map of 
Mrk~370; {\em right panel}: R-band image. North is at the top and East is on 
the left; axis units are arcsec.
\label{Fig:mapas}}
\end{figure}

In the central part of the galaxy are two large star-forming regions (labeled
{\sc 11} and {\sc 12} in Figure~\ref{Fig:mkn370sli}), $\approx 7$ arcsec 
(450~pc) apart, which appear to be connected by a faint bridge-like structure;
these two star-forming regions have been previously cataloged as 
"double nucleus" by \cite{MazBor93}. 
Smaller star-forming knots extend out in the north-east and south-west 
directions, to approximately 40 arcsec (2.5~kpc) from the optical center 
(\ie, the center of the outer isophotes). The star-formation activity is 
concentrated along this direction, with the exception of region {\sc 11}).

We have used the {\sc focas} package to identify and analyze the different 
knots in the continuum-subtracted \Ha\ images. The {\sc focas} task looks for
local maxima and minima in the {\sc adu} counts of each pixel in the image.
We have adopted the criteria that, for a knot to be detected, its area must 
be larger than 12 pixels (corresponding to an equivalent diameter of 2 arcsec 
or 125 pc), to ensure that the diameter of the knot is larger than the 
{\sc psf}, and the counts of each of its pixels must be higher than 3 times 
the standard deviation of the sky background. 
Using these criteria, a total of 16 knots were detected; they are labeled in 
Figure~\ref{Fig:mkn370sli}.\footnote{The Integral continuum maps show another 
fainter knot, not detected by Focas as it does not satisfy these criteria.}

Data on the \Ha\ flux, luminosity, number of ionizing photons, ionized
hydrogen mass and equivalent width of each knot were calculated; these results
are presented in Table~\ref{Tab:hafot}. The number of ionizing photons and the
ionized hydrogen mass were calculated following \citet{kennicutt88}. The
electron temperature and density needed to calculate the mass of the hydrogen 
ionized gas and the number of ionizing photons were obtained from the
long-slit spectra. 

The resulting parameters have been corrected for interstellar extinction and
for the contribution of the [\ion{N}{2}] lines. For the Knots 11 and 12, we
used the spectroscopic parameters derived for the  subregions {\sc b} and {\sc
a} (see Figure~\ref{Fig:mkn370sli}). We used the bi-dimensional data to correct
knots inside the {\sc integral} field. For those knots outside the region
covered by our 2-D spectra, the above corrections have been applied by using
averaged values derived from the  central region.

The luminosity of the brightest knots ({\sc 11} and {\sc 12}) are comparable
to those of individual giant \ion{H}{2} regions 
(typically 10$^{39}$--10$^{40}$ erg sec$^{-1}$; \citealp{Kennicutt91}). 
The range in luminosities is similar to that found for the \ion{H}{2} regions 
cataloged in the Magellanic Irregular NGC 4449 \citep{fuentes00}.

The equivalent width $W(\Ha)$ was computed as:
\begin{equation}
W(\Ha) = C_{\Ha}/C_{\rm cont} \times \Delta W_{\Ha}
\end{equation}
\noindent
where $C_{\Ha}$ is the total number of net counts of the knot, $C_{\rm cont}$ 
is the corresponding total number of counts in the rescaled continuum image, 
and $\Delta W_{\Ha}$ is the width of the \Ha\ filter transmission curve.

The \Ha\ equivalent width has been corrected for the contribution of the
underlying stellar emission to the continuum level; it means that $C_{\rm
cont}$ are the total counts measured in the continuum band after subtracting
the counts due to the emission of the older stars. In order to estimate this 
latter contribution, we assumed that the underlying population can be
described by an exponential model (this procedure is explained in the next
section).
As the exposures in the continuum narrow-band filter were not deep enough to
derive the properties of the underlying disk, the structural parameters
derived in $R$ band have been used.
The continuum flux map through the \Ha\ filter has been derived from the $R$ 
frame after the latter has been rescaled so as to match the continuum in the 
former. We followed the same procedure, based on field stars, commonly 
employed to match \Ha\ and its adjacent continuum (see Paper~II for a 
description of this technique).

In Table~\ref{Tab:hafot} we show the values of W(\Ha) of the different 
star-forming regions in Mrk~370; we have listed the values before and after
applying the correction for the contribution of the older stars, to provide a 
direct estimation of the importance of such contribution. We found the 
difference to be significant in most cases.

\subsection{Broad-band Photometry}

\subsubsection{The Starburst}
\label{Sect:corr} 

The young starburst regions have been isolated in each broad-band filter 
image. Using {\sc focas}, we built a mask where the emission regions 
delimited in the \Ha\ frame have the constant value one and the rest are set 
to zero.
Each broad-band image is then multiplied by this mask. Finally we perform 
aperture photometry on these images, to derive magnitudes and colors of each 
knot. 

Table~\ref{Tab:mkn370bb} shows the broad-band photometry of the sixteen knots
detected in Mrk~370; the measurements have been corrected from Galactic 
extinction using the data published by \cite{BurstHeil84}. 
The extinction coefficient in the other bands were related to the B value 
following \cite{RiekLebo85}.

In order to characterize the stellar population of the individual regions, we
compare their integrated properties with the predictions of evolutionary
synthesis models. This task is especially difficult in this case, because
at least three different contributors are producing the observed light: 
the young stars, the gaseous nebula surrounding them, and the older stars 
underlying the ongoing burst. Broad-band colors must be corrected from the 
contribution of these different factors before they can be compared with 
the model predictions. In order to derive the "true" colors of the starburst, 
we proceeded in the following way.

First, colors were corrected for interstellar reddening, using the value of 
$C(\Hb)$ derived from our long slit spectra. 
For those knots lacking spectroscopic data, we used the $C(\Hb)$ averaged 
over the whole long-slit spectrum (which is probably our best possible guess).

Second, we have corrected for the contribution of emission lines. In gas rich
objects, like BCDs, the presence of strong emission lines in the  spectra can
severely affect broad-band colors; the exact amount of the  contribution of the
lines to the flux observed in a given filter depends  on the equivalent width
of the line and on its precise location under the  filter transmission profile.
We used our spectroscopic information to compute the fraction of flux in the
observed broad-band filters accounted for by each emission line, by using the 
task {\sl calcphot} in the {\sl synphot} package of STSDAS.

Last, we subtracted the emission of the underlying population of stars.
We assumed that the galaxy is composed of the starburst regions delimited by 
the {\sc focas} task, and of an underlying low surface brightness component, 
which is described by an exponential function (see next section). 

The integrated magnitude of a knot, in a given band, $\lambda$, is given by: 

\begin{equation}
\label{totallu}
m_{\lambda}=-2.5 \times \log \left(C_{\lambda}^{SB}+C_{\lambda}^{D}\right) 
+ K_{\lambda}
\end{equation}

\noindent where $C_{\lambda}^{SB}$ is the number of counts in the knot coming 
from the starburst; $C_{\lambda}^{D}$ is the contribution of the underlying 
disk within the knot region, and $K_{\lambda}$ is the corresponding 
calibration constant (which incorporates exposure time, airmass, etc.).

$C_{\lambda}^{D}$ is easily computed from the central surface brightness and 
scale length of the best-fit exponential model, and subtracted out in the 
previous formula, yielding the corrected magnitude of each knot: 

\begin{equation}
m_{\lambda}(SB)=-2.5 \times \log \left(C_{\lambda}^{SB}\right) + K_{\lambda}
\end{equation}

This process has been applied to the $B$, $V$, $R$ and $I$ bands.
The images in $U$ were not deep enough to allow a reliable fit to the light 
profiles; however, by using typical values of \ub\ for an evolved stellar 
population \citep{bruzual93,vazdekis96}, we have estimated the contribution of 
the host galaxy in the $U$ band. 
We found that, in this pass-band, the older component accounts for only 
$\sim 1\%$ of the knot flux, so we can neglect this correction.
The resulting corrected colors are listed in Table~\ref{Tab:mkn370bb}. 

We would like to stress that uncorrected and corrected knot colors differ
significantly. In the case of the Mrk~370,  while emission lines do
not change much the broad band colors (shifts in  \bv\ are $\approx 0.05$ or
less), the internal extinction may change  \bv\ by up to 0.25 mag, and the
effect of the contribution of the LSB stellar component is equally important.
The final color shift is clearly function of the position in the galaxy; this
means that it is dangerous to interpret color gradients as genuine variations
of the knot properties (for instance their age) if the above  corrections have
not been applied.


\subsubsection{The Underlying Stellar Population}

Surface brightness profiles (SBP) in $BVRI$, as well as color profiles of 
Mrk~370, have been presented in Paper~I. The SBP of the galaxy was used to
decompose the galaxy in two components: the starburst, \Ha\ emitting regions,
plus an older LSB stellar host. By fitting an exponential function to the
faint, outer parts of the galaxy light profile (where we expect no or
only negligible contribution from the starburst), we determined the structural
parameters (scale length $\alpha$ and central surface brightness $\mu_0$) of
the LSB host. These parameters were used to calculate its total magnitudes in
the different  bands.

The structural parameters in the different bands, together with the total 
luminosities of the starburst and of the old stellar component, are given in 
Table~\ref{Tab:mkn370under}.

Colors of the underlying component were derived from the best-fit exponential
model; we found $(\bv)_{\rm host} = 0.80$,  $(\vr)_{\rm host} = 0.50$ and
$(\vi)_{\rm host} = 1.46$.

\section{Evolutionary properties of Mrk 370}

\subsection{The Starburst}

In order to constrain the properties of the individual \ion{H}{2} regions, we
compare their observed properties with the predictions of {\em evolutionary 
population synthesis models}. These models assume that stars are born with 
masses distributed according to the initial mass function ({\sc imf}), with a 
star formation rate ({\sc sfr}). The stars are evolved  according to
theoretical evolutionary tracks and finally, using empirical or theoretical
calibrations, the predicted colors or spectra of the composite  population are
obtained. In the last years, people have been doing a lot of work in
evolutionary  synthesis models, see for instance \cite{leitherer96}. Depending
on the  hypothesis on which the different models are based, the predicted
values for  the observables can be slightly different (for example, theoretical
modeling  of the evolution of massive star is still undergoing significant
changes,  \citealp{maeder94}; and revision of these models often leads to new 
generations of synthesis calculations). This means that it is important to
examine the characteristics of the  available models and select the one most
suitable to each specific case.

We have compared the derived observables of the starburst with the predictions
of \citet[hereafter SB99]{leitherer99} evolutionary synthesis models, known as
{\em Starburst99}. These models provide a wide set of predictions for the
spectrophotometric properties of star-forming galaxies. They are an improved
version of the ones previously published by \citet{leitherer95}: the latest set
implements the stellar evolution models of the Geneva group, as well as the
model atmosphere grid compiled by \cite{lejeune97}.

For each of the sixteen delimited knots, we searched for the model which best
matches the observed parameters: W(\Ha), \ub, \bv, \vr\ and \vi. In order to
break the age and metallicity degeneration, we set the metallicity equal to
that derived from the emission line fluxes (which is a reasonable approximation
to the metallicity of a young population). 

We found that, for all the knots, we can reproduce the observed quantities with
an instantaneous burst (IB) of star formation and the Salpeter {\sc imf} with 
an upper mass  limit of 100 $M_{\odot}$. No restriction can be put on the lower
mass, given its poor sensitivity to the young burst evolution, which is
dominated by the  contribution of massive stars ($M> 8 M_{\odot}$). The derived
ages from \ub\  and \Ha\ are quoted in Table~\ref{m370edad}; both observables
give usually  ages consistent with each other.

The ages derived using \bv, \vr\ and \vi\ are slightly larger, but in
general compatible with the former. We notice that these colors are not very
sensitive to the evolution of the young stars, as they vary by about 0.3 to 0.4
mag when the age increases from 0 to 100 Myr (SB99).  Also, these colors are
strongly contaminated by the different factors described in 
Section~\ref{Sect:corr} (reddening, contribution of emission lines and 
emission of the older stars).

Although we have attempted to correct for such factors, the results might be
imprecise due to the various hypotheses we made to work them out.
First, we adopted a constant $C(\Hb)$ in those zone having no spectral
information, while there may be spatial variations in the dust distribution,
and $C(\Hb)$ can vary from knot to knot. That this is likely the case is shown 
by the results we obtained for the two central knots:
$C(\Hb)_{\rm A} = 0.31$ and $C(\Hb)_{\rm B} = 0.39$, which translates
into $E(B-V)_{\rm A} =0.22$, and $E(B-V)_{\rm B} = 0.27$, and $E(V-I)_{\rm
A}=0.34$ and $E(V-I)_{\rm B} = 0.43$.  
If such variations are detected in two knots so close to each other, we can 
expect larger differences for the more peripheral knots. 
The same result is obtained from 2D spectroscopy: the 2D distribution of
$C(\Hb)$ do not present an homogeneous morphology (see Figure 4), but larger
$C(\Hb)$ close to the center of the star-forming knots.
 
The same applies to the contribution of emission lines, though, as mentioned 
earlier, they play a secondary role.

Furthermore, the uncertainties in the structural parameters obtained by the
exponential fit of the LSB component also translate into uncertainties in the
resultant colors. 

All in all, the combined errors on the knot colors due to the various factors 
here discussed can be easily as large as $0.2-0.3$ mag; this fact, together 
with their low sensitivity to the starburst properties, imply that \bv, 
\vr\ and \vi\ alone can not constrain the properties of the young stellar 
population.

On the other hand, \ub\ and \Ha\ are much more sensitive to the evolution of 
the young stars. They drastically change during the first million years of the 
stars life (see Figures 55, 56, 83 and 84 in SB99), which converts them in 
reliable indicators of the properties of star-forming regions. 

\begin{figure}[H]
\centerline{\psfig{figure=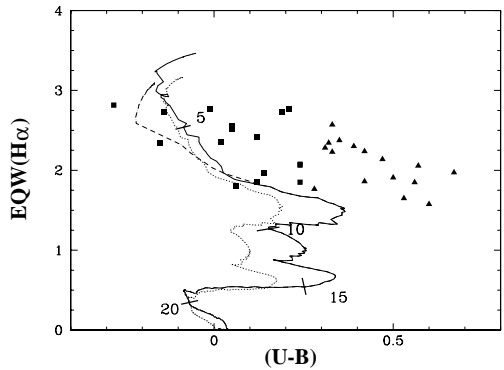,width=7.0truecm}}  
\caption{Equivalent width of the \Ha\ line {\em versus} $(U-B)$.
We also plot the tracks of an instantaneous burst of $z=0.008$, aged from
1 to 30 Myr (Leitherer et al. 1999), for three {\sc imf}s: solid line, 
$\alpha$=2.35, M$_{\rm up}$=100M$_{\odot}$; dotted line: $\alpha$=3.30, 
M$_{\rm up}$=100M$_{\odot}$; short-dash line, $\alpha$=2.35,
M$_{\rm up}$=30M$_{\odot}$.  
Triangles represent the observed (uncorrected) parameters of the individual 
knots; squares represent the knots parameters after applying the corrections 
from reddening and from the contribution of emission lines and older stars.}
\label{Fig:modelo1}
\end{figure}

\begin{figure}[H]
\centerline{\psfig{figure=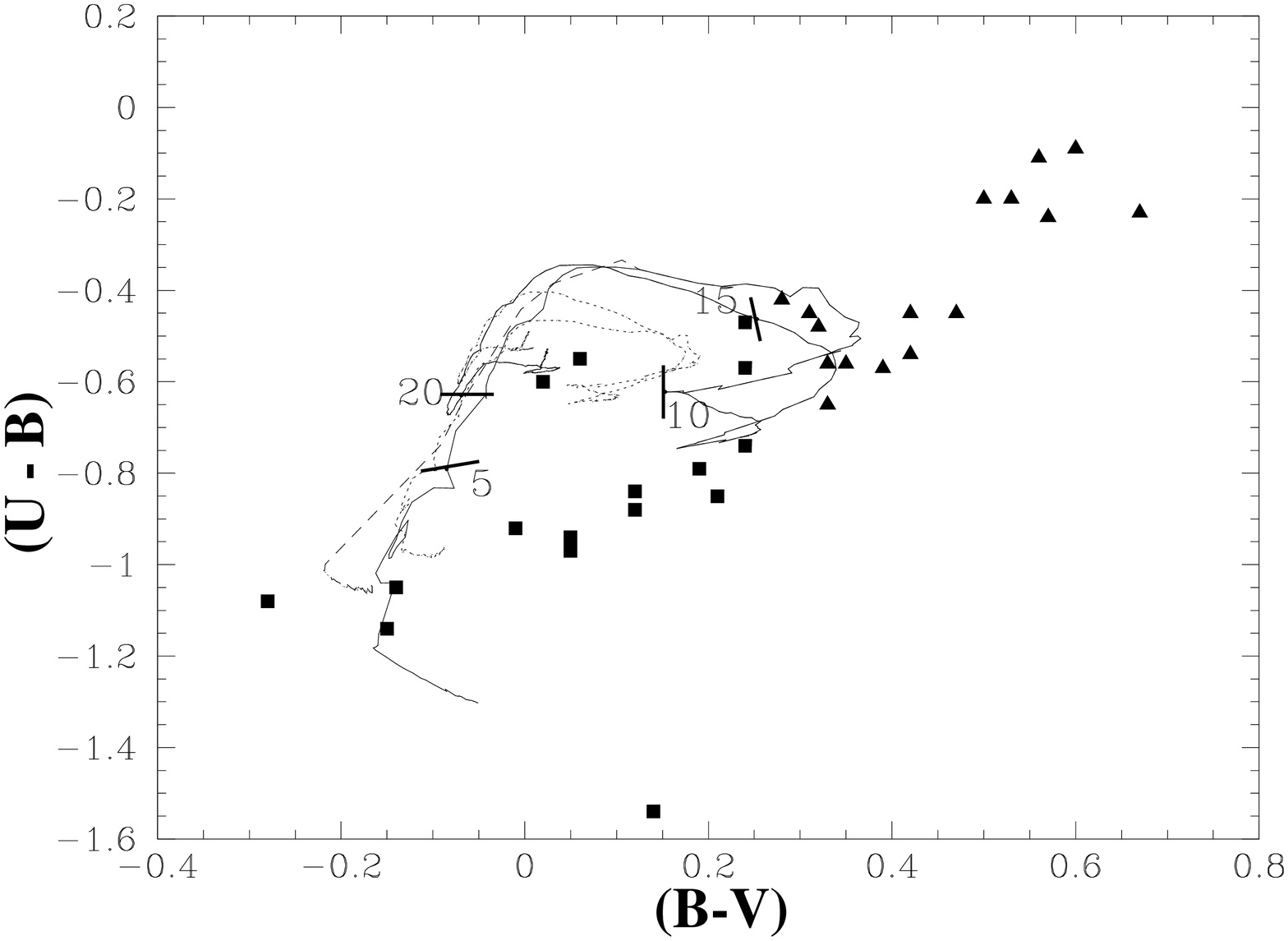,width=7.0truecm}}  
\caption{$(U-B)$ {\em versus} $(B-V)$.
The track of an instantaneous burst of $z=0.008$, aged from
1 to 30 Myr (Leitherer et al. 1999), are also plotted, for three different 
{\sc imf}s: solid line, $\alpha$=2.35, M$_{up}$=100M$_{\odot}$; 
dotted line: $\alpha$=3.30, M$_{up}$=100M$_{\odot}$; 
short-dash line, $\alpha$=2.35,
M$_{up}$=30M$_{\odot}$. Symbols are the same as in the previous figure}
\label{Fig:modelo2}
\end{figure}

We remark that none of the observable parameters will constrain accurately 
the properties of a star-forming region, unless we correct properly for the 
factors described above. The parameters measured directly (\ie without
correction) usually do not fit the models, and, if they do, they tend to give
much older ages. This point is illustrated in  Figures~\ref{Fig:modelo1} and
\ref{Fig:modelo2}.

\subsection{Age of the older stars}

The \vi\ color appears to be much too red to be compatible with a normal older
stellar population. The data in the $I$ band are not as good and deep as in the
other bands, and the large uncertainties on the $I$-band  structural parameters
may be responsible for this discrepancy. Thus, the final ages have been derived
by using the \bv\ and \vr\ colors. These colors have not been corrected from
interstellar extinction, because the extinction coefficients we derived were
based on the gas emission lines, and they may not apply to the regions outside
the area occupied by the starburst. We have assumed that the metallicity of the
stellar population is lower than the metallicity derived for the gas.

We have compared the \bv\ and  \vr\ colors derived for the LSB component with
the predictions of two different groups of evolutionary synthesis models: 

{\em The Galaxy Isochrone Synthesis Spectral Evolutionary Library, GISSEL~96}
\citep{leitherer96,BruzChar93}, which include simple stellar populations (SSP)
or IB with different metallicities from $Z=0.0004$ to $Z=0.10$. The age varies
from 0 to 20 Gyr, and Salpeter and Scalo {\sc imf}s are considered. We have
also used the previous version of the models (GISSEL~95) in order to include a
constant star forming rate (CSFR) ---~only available for solar metallicity.

The {\em Spectrophotometric Population Synthesis Library for Old Stellar
Systems}  \citep{vazdekis96}, a set of stellar population synthesis models
designed to study old stellar systems. They synthesize SSP or follow the
galaxy through its evolution from an initial gas cloud to the present time,
assuming an analytical functional form for the SFR (see Section 3.1 in
\citealp{vazdekis96}), and including the chemical evolution. 
Two different {\sc imf}s are considered: a unimodal {\sc imf}, with a power 
law form whose slope is a free parameter, and a bimodal {\sc imf}, equal to 
the unimodal {\sc imf} for stars with masses above 0.6 $M_{\odot}$, but with 
a reduced influence of the stars with lower masses.  

When comparing with the predictions of the IB approximation, both models 
give ages of $\sim 5$ Gyr for the older stellar component of Mrk~370. 

However, when changing the star forming history, we obtained different results.
the \bv\ and  \vr\ colors could not be reproduced by GISSEL95 with a CSFR.
Next, we applied \cite{vazdekis96} models, which include chemical evolution. We
considered  Salpeter {\sc imf}, and the different  $\nu$ ($\nu$ is a constant
fixing the timescale of star formation, see Sect 3.1 in \citealp{vazdekis96}),
provided by the models. We found the following results: when $\nu=1$ the
observed colors are inconsistent with the predictions of the  models; with
$\nu=5$ colors agree with ages between 9 and 11 Gyr; and in the  rest of the
cases ($\nu=10,20$ and $50$) colors are consistent with ages  between 4 and 7
Gyrs.  

Optical colors are not very sensitive to the evolution of the older stars; for
instance, \bv\ increase by $\approx$ 0.1 mag when ages varies from 5 to 17 Gyr 
and the same applies to \vr. Such variations are of the order of the 
uncertainties in our measured colors.  
Although no firm constraints could be put on the age of the underlying stellar 
population, the above comparison with models point to a minimum age of 5 Gyrs; 
but it is clear that in order to better constrain the age of the LSB host, it 
is indispensable to include near-infrared colors in the analysis.

\section {Discussion}

It turns out from the above results that the kind of spectrophotometric study
we have carried out is fundamental in order to disentangle the stellar 
populations in BCDs and derive their star forming histories. 
However, as we have already pointed out in the introduction, most of the work 
carried out so far has focused on statistical analysis of BCDs samples, 
while only a few papers have been devoted to examining in detail the 
characteristic of individual objects: 

The two extremely compact and low-metallicity BCDs, SBS~0335-052 and 
Tololo~65, both classified as i0 (LT86), have been the subject of recent 
spectrophotometric studies by \citet{polis98,polis99}.
They found in both cases that the properties of the underlying component are 
consistent with those of a stellar population not older than one Gyr.

\citet{noeske00} report on results from a spectrophotometric analysis of  two
iI,C ("Cometary BCDs", LT86) Mrk~59 and Mrk~71. Spectral population synthesis
models, in combination with color magnitudes diagrams and color profiles, yield
a most probable formation age of $\approx$ 2 Gyr for the old stars in both
galaxies.  Two other cometary BCDs have been studied by  \citet{fricke00} 
---~Tololo~1214-277~--- and \citet{guseva01} ---~SBS0940+544. In the latter
case, no compelling evidences favor either a young or an old age.  

Despite the fact that nE/iEs are the most common type of BCD, there are very 
few studies, with the same quality, of objects belonging to this class.
\citet{steel96} and \citet{mendez99} have performed spectrophotometric
observations of the iE BCDs Haro~3 and II~Zw~33 respectively. Although in both
cases an underlying component of older stars was detected, their data did not
permit to determine its properties, and both studies limited the analysis to
the young starburst. The ages and star-forming scenarios they found for these
two objects are consistent with our findings. They also found metallicities
comparable to the metallicity we have derived for Mrk~370. More recently, a
comprehensive analysis of the iE BCD galaxy Mrk~86 has been published by
\citet{gildepaz00a} and  \citet{gildepaz00b}. They studied the properties of 46
individual star-forming knots as well as the underlying population of older
stars. By applying evolutionary synthesis models, they found that three well
defined stellar populations are present in Mrk~86:  the star-forming regions,
with ages between 5 and 13 Myr, and no significant age or metallicity gradient,
a central starburst with an intermediate age of 30 Myr, and an underlying
population of stars, with a surface brightness profile that can be described by
an exponential, with no color gradients and an age between 5 and 13 Gyrs. 

In all the above studies the star-forming knots present a similar range of
ages, consistent with our findings for Mrk~370. However, the galaxies appear
to be different with respect to the properties of their underlying population 
of stars: while i0 and cometary BCDs seem to be less evolved systems, 
the structural properties and ages derived for the LSB of Mrk~370 are quite
similar to those presented by the iE BCD Mrk~86. This fact may imply that 
BCDs are objects which have in common the presence of an active starburst, 
which dominates their optical properties but, when the characteristics of the
underlying stellar component are taken into account, they form a more
heterogeneous class. This picture could naturally explain the different
evolutionary stages derived for the galaxies.

We argue that considering BCD as a class might be then inappropriate.
Therefore, looking for absolute unifying scenarios ---~for instance, trying to
link the whole BCD class with the different types of dwarf galaxies (when only
a fraction of them could be actually connected)~--- may be speculative at this
time.

On the other hand, \cite{noeske00} suggested that an evolutionary sequence 
might exist connecting the different BCD subtypes, the iI,C galaxies being a 
possible link between the extremely young galaxy candidates (iO class) and the 
more evolved iE/nE BCDs. 

The information available at the present time does not permit to discard any of
these hypothesis. Comprehensive studies of larger samples of BCDs are 
required in order to find out which scenario is the correct one.

Our kinematic results suggest ordered motion around the north-south axis of
Mrk~370, the kinematic center being located close to the center of the external
isophotes; this could imply that the gas is rotating coupled with the main
body of the galaxy (the underlying population hosting the starburst).
However, we should take into account that, because of our limited spectral
resolution, more complex gas motions could escape detection. In fact,
\citet[2001]{oestlin99}, analyzing high-resolution (velocity sampling 
$\simeq 5$ km sec$^{-1}$) Fabry-Perot observations of BCD galaxies, found 
that, in general, the velocity fields of the studied galaxies are irregular 
and distorted. 
Spectroscopic observations of Mrk~370 with high velocity resolution are needed 
to better trace its gas kinematics.

Besides, Mrk~370 present some other puzzling peculiarities: the peak in the
continuum map does not coincide with the kinematic center, nor with the center
of the outer isophotes, nor with one of the knots; and the position angle of
the outer regions, seen in the $R$-band grey-scale map, differs by about 30
degrees from the position angle displayed by the \Ha\ map.

\section{Summary and Conclusions}

Broad- and narrow- band images of the BCD Mrk~370, together with long-slit and 
two-dimensional spectra, have been analyzed in order to derive the properties 
of the different components of the galaxy and to constrain its evolutionary 
status. Our results can be summarized as follow:

\begin{itemize}

\item Two different stellar components are clearly distinguished in the galaxy:
the present starburst and an underlying older population. The current star
formation activity takes place in numerous knots, aligned in a north-east 
south-west direction. An extended LSB component, with regular appearance, 
hosts this star-forming area.

\item The oxygen abundance derived from the spectra is $Z \simeq Z_{\odot}/2$, 
a value relatively high for the BCD class; the high blue continuum, the 
absorption wings in the Balmer lines and the pronounced Balmer discontinuity 
in its spectrum are clear indicators of the presence of an evolved stellar 
population.

\item The starburst region is resolved into small and compact star-forming 
regions. We identified a total of sixteen knots, and derived their individual 
properties. 
The brightest knots have luminosities similar to those of giant \ion{H}{2}
regions in M33 and in the LMC.

The photometric parameters of the individual star-forming knots were corrected 
from the contribution of emission lines, internal extinction and emission from 
the underlying LSB component. The colors before and after the corrections are 
significantly different. In Mrk~370 the internal extinction and the 
contribution of older stars are more important than emission lines in shifting
the observed broad-band colors of the knots
The color shifts are function of the position of the knot in the galaxy, and 
we strongly caution against interpreting observed color gradients as age 
gradients {\em before\/} applying such corrections.

We found that we can reproduce the colors of all the knots with an IB of star 
formation and the Salpeter initial mass function ({\sc imf}) with an upper 
mass limit of 100 M$_{\odot}$. 
The ages of the knots range between 3 and 6 Myrs.

\item The age of the underlying stellar population has been estimated by
comparing its optical colors with two different groups of evolutionary 
synthesis models. Colors are consistent with ages larger than 5 Gyr.    

\item The inner, brightest star forming-regions, previously cataloged as 
double nucleus, are normal \ion{H}{2} regions. 
We would like to stress that the "multiple nuclei" often seen in broad-band 
images of BCD galaxies may be just the superposition of off-center luminous 
\ion{H}{2} regions. If Mrk~370 is not a double-nucleus galaxy, then there may 
be no compelling evidence that it has experienced recent interactions.
Indeed, there are no other signs of interactions; the regularity of the 
velocity field, and the fact that the central knots share the same kinematics 
of the main body point against a recent merger or interaction event. 
Also, \cite{Noeskeetal01} report that Mrk~370 is an isolated galaxy.

\item We obtained the mean ionized gas velocity field in the central part of
the galaxy. The velocity field seems to be regular and similar to that of a
rotating disk (west region approaching, east region receding).

\end{itemize}

All in all, spectrophotometric studies of BCDs are of paramount importance in
order to  derive their evolutionary status. These analyses must include, in
addition to  the whole optical dataset, near-infrared photometry (NIR)
---~colors in the NIR better trace the  properties of the older stellar
component~--- as well as deep optical spectrophotometry of the host galaxy. 
Besides, high resolution spectroscopy of BCDs will help trace the kinematics of
the galaxy and provide  valuable information on the mechanisms triggering the
star formation.

\acknowledgments

Based on observations with the JKT, INT  and WHT, operated on the island of 
La Palma by the Royal Greenwich Observatory in the Spanish Observatorio 
del Roque de los Muchachos of the Instituto de Astrof\'\i sica de Canarias. 
Based also on observations taken at the German-Spanish Astronomical Center, 
Calar Alto, Spain, operated by the Max-Planck-Institut f{\" u}r Astronomie 
(MPIA), Heidelberg, jointly with the spanish "Comision Nacional de Astronomia".
We thank the staff of both observatories.
This research has made use of the NASA/IPAC Extragalactic Database (NED), 
which is operated by the Jet Propulsion Laboratory, Caltech, under contract 
with the National Aeronautics and Space Administration. 

We thank J. Iglesias-P\'aramo and J.N. Gonz\'alez-P\'erez for their help in 
the initial stages of this project. 
We thank A. Vazdekis, P. Papaderos and K. Noeske for valuable comments and 
discussions. 
We also acknowledge the anonymous referee for his/her helpful comments which 
helped us improve this paper.

This work has been partially funded by the spanish ``Ministerio de Ciencia
y Tecnologia'' (grants AYA2001-3939 and PB97-0158).
L.M. Cair\'os acknowledges support by the EC grant HPMF-CT-2000-00774.

\clearpage

\begin{deluxetable}{lccccccc}
\footnotesize
\tablewidth{0pt}
\tablecaption{Basic data of Mrk~370}
\tablehead{
  \colhead{Galaxy}   & \colhead {Other designations} & \multicolumn{2}{c}{R.A.\ \ (1950) \ \ Dec.} & 
  \colhead{$M_{B}$}  & \colhead{D (Mpc)} & \colhead{$M_{HI}^{1}$} &
  \colhead{$M_{T}^{1}$}}
\startdata
 Mrk~370     & NGC~1036,  & $02^{\rm h}$ $37^{\rm m}$ $40^{\rm s}$
             & $19^\circ$ $05'$ $01''$ &  $-17.20$ & 12.9 &
             0.36$\times$10$^{9}$ & $4.2\times 10^{9}$ \\
             & UGC~02160  &  &  & & & & \\ 
\enddata
\label{Tab:Galcoord}
\tablecomments{(1) Neutral hydrogen mass $M_{HI}$ and total mass $M_{T}$ in
units of $M_\odot$; both from Thuan and Martin (1981).}
\end{deluxetable}

\newpage

\begin{deluxetable}{ccccc}
\tablewidth{0pt}
\tablecaption{Log of the observations \label{Tab:Obslog}}
\tablehead{
\colhead{Date} & \colhead{Telescope} & 
            \colhead{Instrument} & \colhead{Filter/grism} 
            & \colhead{Exposure time (s)} }
\startdata
         Oct. 98 & JKT 1.0m       & Cass. focus & U               & 2400 \\
         Oct. 98 & JKT 1.0m       & Cass. focus & B               & 1500\\
         Oct. 98 & JKT 1.0m       & Cass. focus & V               & 1000\\
         Oct. 98 & JKT 1.0m       & Cass. focus & R               & \phn 800 \\
         Oct. 98 & JKT 1.0m       & Cass. focus & I               & 1200\\
         Dec. 97 & CAHA 2.2m      & CAFOS       & 6569 (113)      & 5400 \\
         Dec. 97 & CAHA 2.2m      & CAFOS       & 6462 ( 98)      & 5400 \\
         Sep. 98 & INT 2.5m       & IDS         & R300V           & 1800 \\
         Aug. 99 & WHT 4.2m       & INTEGRAL    & 600g/mm         & 3600\\[4pt]
\enddata
\tablecomments{
JKT = Jacobous Kapteyn Telescope, ORM (La Palma), Spain (detector:
Tek $1024\times1024$, $0\farcs33$/pixel).
CAHA = Centro Astron{\'o}mico Hispano Alem{\'a}n, Almer{\'\i}a, Spain
(Site-1D $2048\times2048$, $0\farcs53$/pixel)/
INT = Isaac Newton Telescope, ORM (La Palma), Spain 
(EEV10 $4100\times2048$ $13.5\mu$ pixels).
WHT = William Herschel Telescope, ORM (La Palma), Spain
(Tex $1024\times1024$ $24\mu$ pixels).}
\end{deluxetable}

\begin{deluxetable}{ccccccccc}
\rotate
\tablewidth{0pt}
\tablecaption{Reddening corrected line intensity ratios}
\tablehead{\colhead{Line}  &  \colhead{Ion}     & \colhead{$f(\lambda)$}  & 
\multicolumn{2}{c}{A} & \multicolumn{2}{c}{B}  & \multicolumn{2}{c}{S} \\
\colhead{$(\AA)$}            &            &           & 
\colhead{$F_{\lambda}$}   & \colhead{$-W_{\lambda}$}  & 
\colhead{$F_{\lambda}$}   & \colhead{$-W_{\lambda}$}  &  
\colhead{$F_{\lambda}$}   & \colhead{$-W_{\lambda}$}}
\startdata
3727& [\ion{O}{2}]  & \phs0.26       &$2.74\pm0.06$          &   $35.1\pm0.9$
                                     &$3.22\pm0.09$          &   $136\phd\phn\pm\phn10\phd\phn$
                                     &$2.68\phn\pm0.04\phn$  &   $48\phd\phn\pm1\phd\phn$  \\
3869& [\ion{Ne}{3}] & \phs0.23       &$0.50\pm0.07$          &   $\phn5.1\pm0.8$
                                     &$1.1\phn\pm0.2\phn$    &   $330\phd\phn\pm166\phd\phn$
                                     &$0.49\phn\pm0.06\phn$  &   $\phn7\phd\phn\pm1\phd\phn$  \\
4101& H$\delta$  & \phs0.18          &$0.28\pm0.03$          &   $\phn2.1\pm0.4$
                                     &$0.24\pm0.04$          &   $\phn\phn7\phd\phn\pm\phn\phn1\phd\phn$
                                     &$0.26\phn\pm0.02\phn$  &   $\phn2.5\pm0.3$  \\
4340& H$\gamma$  & \phs0.14          &$0.45\pm0.02$          &   $\phn3.0\pm0.2$
                                     &$0.50\pm0.02$          &   $\phn14.0\pm\phn\phn0.8$
                                     &$0.49\phn\pm0.07\phn$  &   $\phn5\phd\phn\pm1\phd\phn$  \\
4861& \Hb\       & \phs0\phd\phn\phn &$1.00\pm0.01$          &   $12.6\pm0.1$
                                     &$1.00\pm0.02$          &   $\phn26.9\pm\phn\phn0.8$
                                     &$1.00\phn\pm0.01\phn$  &   $16.0\pm0.3$ \\
4959& [\ion{O}{3}] & $-0.02$         &$0.53\pm0.01$          &   $\phn6.1\pm0.2$
                                     &$0.60\pm0.02$          &   $\phn15.5\pm\phn\phn0.6$
                                     &$0.52\phn\pm0.01\phn$  &   $\phn7.6\pm0.1$\\
5007& [\ion{O}{3}]  & $-0.03$        &$1.60\pm0.02$          &   $19.2\pm0.2$
                                     &$1.93\pm0.03$          &   $\phn51.2\pm\phn\phn0.6$
                                     &$1.63\phn\pm0.02\phn$  &   $25.0\pm0.2$ \\
5876& HeI        & $-0.23$           &$0.06\pm0.01$          &   $\phn0.8\pm0.1$
                                     &$0.10\pm0.01$          &   $\phn\phn3.5\pm\phn\phn0.3$
                                     &$0.069\pm0.006$        &   $\phn1.3\pm0.1$   \\
6548& [\ion{N}{2}] & $-0.34$         &$0.14\pm0.01$          &   $\phn2.5\pm0.2$
                                     &$0.11\pm0.01$          &   $\phn\phn4.2\pm\phn\phn0.6$
                                     &$0.149\pm0.008$        &   $\phn3.4\pm0.2$  \\
6563& \Ha\       & $-0.34$           &$2.85\pm0.05$          &   $56.8\pm0.6$
                                     &$2.93\pm0.09$          &   $112\phd\phn\pm\phn\phn2\phd\phn$
                                     &$2.86\phn\pm0.03\phn$  &   $68.1\pm0.7$\\
6584& [\ion{N}{2}] & $-0.34$         &$0.50\pm0.02$          &   $10.1\pm0.5$
                                     &$0.49\pm0.02$          &   $\phn24\phd\phn\pm\phn\phn2\phd\phn$
                                     &$0.476\pm0.009$        &   $11.6\pm0.3$\\
6678& \ion{He}{1} & $-0.35$          &$0.04\pm0.02$          &   $\phn0.8\pm0.4$
                                     &$0.06\pm0.03$          &   $\phn\phn3\phd\phn\pm\phn\phn1\phd\phn$
                                     &$0.030\pm0.008$        &   $\phn0.7\pm0.2$ \\
6717& [\ion{S}{2}] & $-0.36$         &$0.50\pm0.01$          &   $10.0\pm0.2$
                                     &$0.45\pm0.02$          &   $\phn21.3\pm\phn\phn0.9$
                                     &$0.471\pm0.008$        &   $11.5\pm0.2$\\
6731& [\ion{S}{2}] & $-0.36$         &$0.38\pm0.01$          &   $\phn7.4\pm0.3$
                                     &$0.34\pm0.02$          &   $\phn17\phd\phn\pm\phn\phn2\phd\phn$
                                     &$0.365\pm0.007$        &   $\phn9.0\pm0.2$    \\
7135& [\ion{Ar}{3}]& $-0.41$         &$0.11\pm0.01$          &   $\phn2.4\pm0.4$
                                     &$0.10\pm0.01$          &   $\phn\phn4.9\pm\phn\phn0.8$
                                     &$0.093\pm0.006$        &   $\phn2.5\pm0.2$  \\[6pt]
$C(\Hb)$   & &         & \multicolumn{2}{c}{ 0.31$\pm$0.02}&  \multicolumn{2}{c}{0.39$\pm$0.03}&   \multicolumn{2}{c}{0.285$\pm$0.004}\\
W$_{\rm abs} (\AA)$ & &      & \multicolumn{2}{c}{1.3}&\multicolumn{2}{c}{0.0} & \multicolumn{2}{c}{1.4}\\
$F(\Hb)$    & &         & \multicolumn{2}{c}{2.67$\pm$0.0014}& \multicolumn{2}{c}{1.55$\pm$0.12}&
\multicolumn{2}{c}{$3.87\pm0.06$}\\      
\enddata
\tablecomments{Reddening-corrected line intensities (normalized to $\Hb=1$) 
for the regions extracted from the long slit spectrum of Mrk~370.
The integrated spectrum S is presented in the rightmost columns.
Balmer lines are corrected from underlying stellar absorption.
The reddening coefficient, $C(\Hb)$, the value of the absorption correction, 
W$_{\rm abs}$, and the measured \Hb\ flux, $F(\Hb)$ ($\times 10^{-14}$ erg 
cm$^{-2}$ sec$^{-1}$) are also included. 
\label{Tab:flux} }
\end{deluxetable}

\begin{deluxetable}{rcccccc}
\tablewidth{0pt}
\tablecaption{\Ha\ photometry of the individual knots in Mrk~370
\label{Tab:hafot}}
\tablehead{
\colhead{Knot} & \colhead{Flux(\Ha)} &
            \colhead{$\log L(\Ha)$} & \colhead{$\log N_{Lym}$} &
            \colhead{$\log(M_{HII}/M_{\odot})$}  &
            \colhead{$-W(\Ha)$} & \colhead{$-W(\Ha)_{\rm corr}$}}
\startdata
1  &  \phn\phn5.70  & 38.05  & 49.92  & 3.13 &     201   &     364 \\
2  &  \phn\phn1.96  & 37.59  & 49.46  & 2.67 &     170   &     534 \\
3  &  \phn\phn5.08  & 38.00  & 49.87  & 3.08 &     138   &     260 \\
4  &  \phn\phn8.01  & 38.20  & 50.01  & 3.28 &     173   &     329 \\
5  &  \phn\phn4.13  & 37.91  & 49.78  & 2.99 &  \phn45   &  \phn71 \\
6  &  \phn\phn0.68  & 37.13  & 49.00  & 2.20 &  \phn94   & \nodata \\
7  &  \phn\phn1.64  & 37.51  & 49.38  & 2.59 &  \phn38   &     116 \\
8  &     \phn37.30  & 38.87  & 50.74  & 3.95 &  \phn58   &  \phn63 \\
9  &  \phn\phn7.29  & 38.16  & 50.03  & 3.24 &  \phn81   &     119 \\
10 &  \phn\phn4.56  & 37.96  & 49.82  & 3.03 &     114   &     578 \\
11 &        147.07  & 39.47  & 51.33  & 4.64 &     373   &     533 \\
12 &        232.17  & 39.66  & 51.53  & 4.96 &     192   &     226 \\
13 &  \phn\phn2.32  & 37.66  & 49.53  & 2.74 &  \phn71   & \nodata \\
14 &     \phn33.31  & 38.82  & 50.69  & 3.90 &     237   &     579 \\
15 &  \phn\phn4.01  & 37.90  & 49.77  & 2.98 &  \phn73   &     223 \\
16 &  \phn\phn5.39  & 38.03  & 49.90  & 3.11 &     220   &     656 \\ 
\enddata
\tablecomments{The data are corrected from interstellar extinction and from
[\ion{N}{2}] emission.
The equivalent widths are shown before and after correction from the
contribution of the underlying stellar emission to the underlying continuum
(except knots 6 and 13).
\Ha\ fluxes are in $10^{-15}$ erg cm$^{-2}$ sec$^{-1}$ units;
\Ha\ luminosities in  erg sec$^{-1}$ units. $W(\Ha)$ in Amstrongs}

\end{deluxetable}

\begin{deluxetable}{rcccccccccc}
\rotate
\tablewidth{0pt}
\tablecaption{Broad-band photometry of the individual knots
detected in Mrk~370. \label{Tab:mkn370bb}}
\tablehead{
\colhead{Knot} & \colhead{$B$} & 
            \colhead{$U-B$} & \colhead{$B-V$} & 
            \colhead{$V-R$} & 
            \colhead{$V-I$} & \colhead{$B_{\rm c}$} &
            \colhead{$(U-B)_{\rm c}$} & \colhead{$(B-V)_{\rm c}$}&
            \colhead{$(V-R)_{\rm c}$} & \colhead{$(V-I)_{\rm c}$}}
\startdata
1     &  20.19 & $-0.57$  &  0.39  & 0.25 &   0.88 &  19.67 & $-0.94$ & \phs0.05     &   \phs0.04 &  \phs0.24\\
2     &  21.01 & $-0.56$  &  0.33  & 0.38 &   1.01 &  20.70 & $-1.05$ &   $-0.14$    &   \phs0.23 &  \phs0.26\\
3     &  20.15 & $-0.45$  &  0.47  & 0.39 &   1.00 &  19.69 & $-0.88$ & \phs0.12     &   \phs0.21 &  \phs0.47\\
4     &  19.88 & $-0.54$  &  0.42  & 0.44 &   1.02 &  19.43 & $-0.97$ & \phs0.05     &   \phs0.30 &  \phs0.47\\
5     &  19.28 & $-0.20$  &  0.53  & 0.38 &   0.96 &  18.76 & $-0.57$ & \phs0.24     &   \phs0.17 &  \phs0.45\\
6     &  21.82 & $-0.23$  &  0.67  & 0.31 &   1.41 &  22.24 & $-1.54$ & \phs0.14     &    $-$0.04 &  \phs1.20\\
7     &  20.11 & $-0.09$  &  0.60  & 0.41 &   1.07 &  19.88 & $-0.74$ & \phs0.24     &   \phs0.17 &  \phs0.46\\
8     &  16.80 & $-0.42$  &  0.28  & 0.29 &   0.66 &  16.04 & $-0.55$ & \phs0.06     &   \phs0.10 &  \phs0.33\\
9     &  19.17 & $-0.20$  &  0.50  & 0.40 &   0.94 &  18.56 & $-0.47$ & \phs0.24     &   \phs0.18 &  \phs0.55\\
10    &  20.00 & $-0.24$  &  0.57  & 0.47 &   1.13 &  19.73 & $-0.85$ & \phs0.21     &   \phs0.24 &  \phs0.66\\
11    &  17.46 & $-0.65$  &  0.33  & 0.39 &   0.69 &  17.37 & $-0.79$ & \phs0.19     &   \phs0.28 &  \phs0.46\\
12    &  15.77 & $-0.45$  &  0.31  & 0.33 &   0.66 &  14.80 & $-0.60$ & \phs0.02     &   \phs0.05 &  \phs0.21\\
13    &  20.13 & $-0.11$  &  0.56  & 0.48 &   1.06 &  19.97 & $-0.84$ & \phs0.12     &   \phs0.10 &  \phs0.43\\
14    &  18.43 & $-0.56$  &  0.35  & 0.41 &   0.82 &  17.90 & $-0.92$ &   $-0.01$    &   \phs0.14 &  \phs0.17\\
15    &  19.81 & $-0.45$  &  0.42  & 0.41 &   1.25 &  19.61 & $-1.14$ &   $-0.15$    &   \phs0.26 &  \phs0.76\\
16    &  20.50 & $-0.48$  &  0.32  & 0.47 &   0.84 &  20.21 & $-1.08$ &   $-0.28$    &   \phs0.45 &   $-$1.68\\
\enddata
\tablecomments{Columns 2 to 6: only the Galactic extinction correction has 
been applied; columns 7 to 11: values corrected from interstellar extinction,  
contribution of emission line and emission from the underlying host galaxy.}
\end{deluxetable}

\begin{deluxetable}{cccccc}
\tablewidth{0pt}
\tablecaption{Structural parameters characterizing the LSB host.
\label{Tab:mkn370under}}
\tablehead{
\colhead{Band} & \colhead{$\mu_{0}$} & 
            \colhead{$\alpha$} & \colhead{$M_{\rm SB}$} & 
            \colhead{$M_{\rm host}$ } & 
            \colhead{$R_{\rm SB/host}$}}
\startdata
$B$    & 22.56   & 1.06  & $-16.69$  & $-16.13$ & 1.68  \\
$V$    & 21.74   & 1.05  & $-16.91$  & $-16.93$ & 0.98  \\
$R$    & 21.04   & 0.96  & $-17.32$  & $-17.44$ & 0.89  \\
$I$    & 20.57   & 1.14  & $-17.23$  & $-18.28$ & 0.38  \\
\enddata
\tablecomments{Columns 2, 3: centrals surface brightness (mag arcsec$^{-2}$) 
and scale-length (kpc); column 3, 4: absolute magnitude of the starburst and of
the host LSB in each band; column 5: the ratio of the luminosity of the
starburst over the luminosity of the underlying component.
As mentioned in the text, the $U$-band image was too shallow to reach out the 
outer, starburst-free regions.}
\end{deluxetable}

\begin{deluxetable}{rcc|rcc}
\tablewidth{0pt}
\tablecaption{Derived ages (in Myr) for the individual star-forming knots in
Mrk~370 \label{m370edad}}
\tablehead{
\colhead{Knot} & \colhead{Age$_{U-B}$} &
    \colhead{Age$_{EW(\Ha)}$} & \colhead{Knot} & \colhead{Age$_{U-B}$} &
    \colhead{Age$_{EW(\Ha)}$}}
\startdata
1     &  3.2--4.8     & 4.8--5.1    &  9        &  5.6--6.4     & 5.9--6.3  \\
2     &  3.1--3.8     & 4.6--4.8    & 10        &  4.5--5.0     & 4.6--4.8  \\
3     &  3.5--5.0     & 5.0--5.5    & 11        &  4.6--5.1     & 4.6--4.8  \\
4     &  3.2--4.7     & 4.8--5.2    & 12        &  5.1--5.9     & 5.1--5.7  \\
5     &  5.1--6.0     & 6.4--7.0    & 13        &  4.5--5.1     & ---      \\
6     &  ---          & 6.1--6.6    & 14        &  3.2--4.8     & 4.6--4.8  \\
7     &  4.8--5.3     & 5.9--6.3    & 15        &  2.2--3.3     & 5.1--5.7  \\
8     &  5.3--6.2     & 6.4--7.0    & 16        &  2.9--3.5     & 4.3--4.7  \\
\enddata
\end{deluxetable}

\clearpage

\clearpage

\clearpage

\clearpage

\section{APPENDIX A}

\subsection{Atlas of Spectra}
\begin{figure}[H]                         
\centerline{\psfig{figure=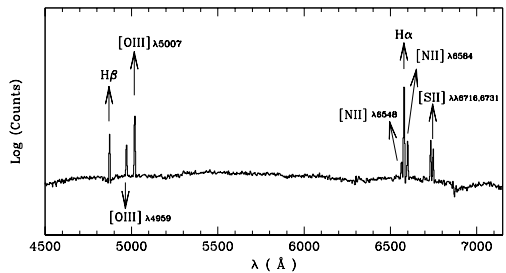,width=7.0truecm}}
\caption{Nuclear spectrum of Mrk~370, obtained by averaging the 7 fibers 
closest to the continuum maximum ($r < 1''.2$).}
\label{apen1}
\end{figure}

\begin{figure}[H]                         
\centerline{\psfig{figure=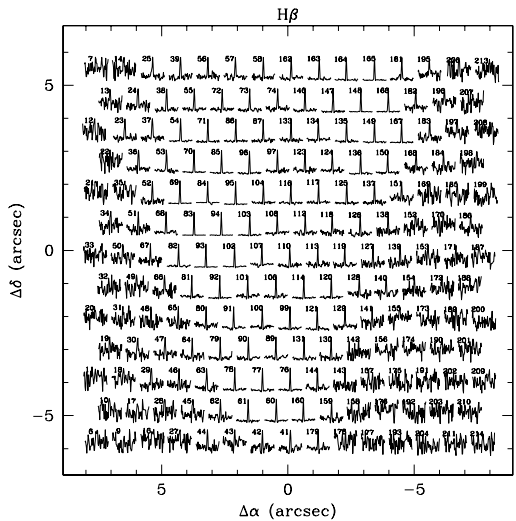,width=7.0truecm}}
\caption{Spectra distribution diagrams of the central $16''\times12''$ of
Mrk~370. Two-dimensional distribution of the \Hb\ lines (the absorption 
wings are visible). 
The spectral range plotted is 4795--4950 \AA.}
\label{apen2}
\end{figure}

\begin{figure}[H]                         
\centerline{\psfig{figure=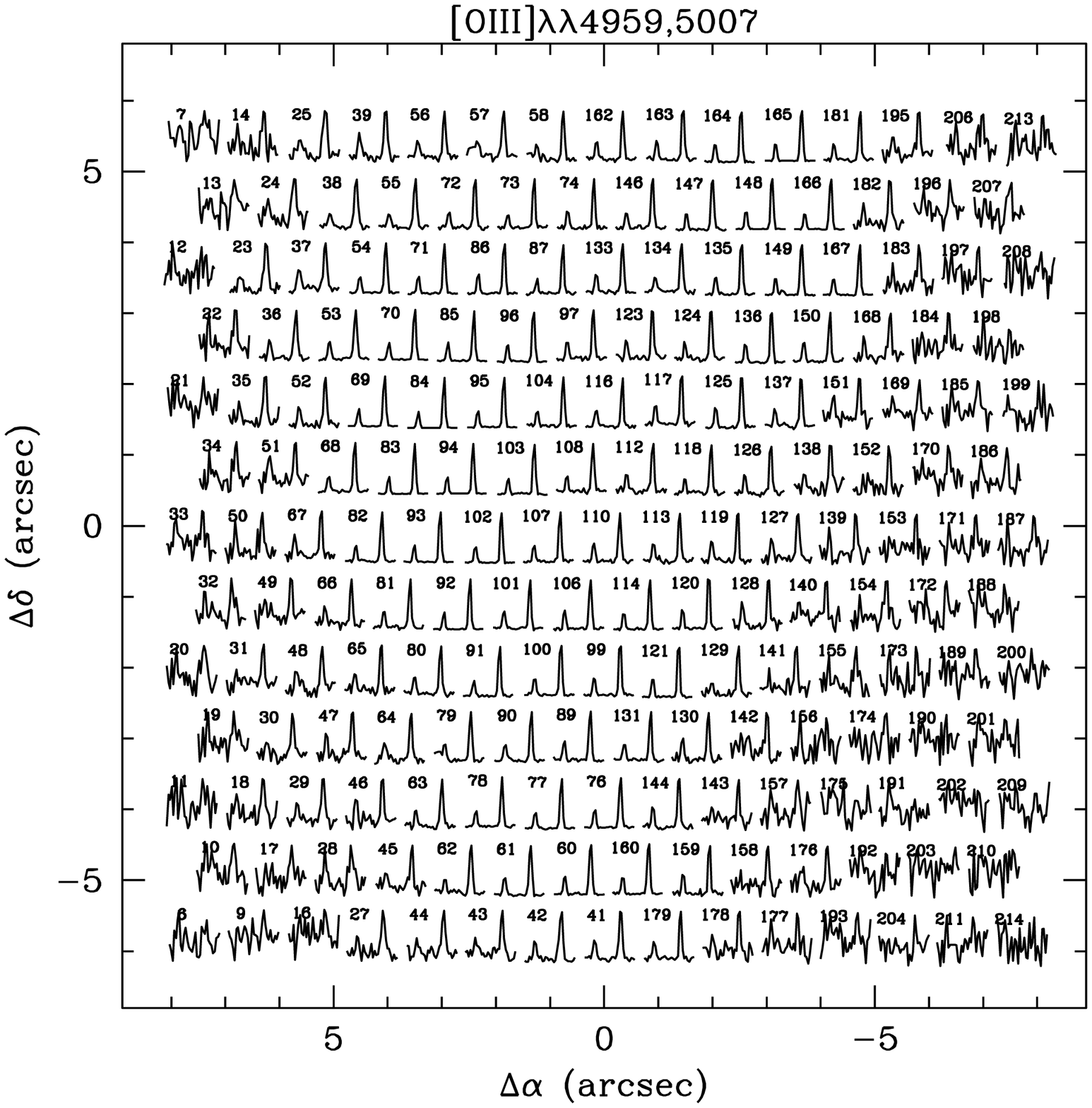,width=7.0truecm}}
\caption{Two-dimensional distribution of 
[\ion{O}{3}]$\lambda4959+$[\ion{O}{3}]$\lambda5007$ (4950--5045 \AA).}
\label{apen3}
\end{figure}

\begin{figure}[H]                         
\centerline{\psfig{figure=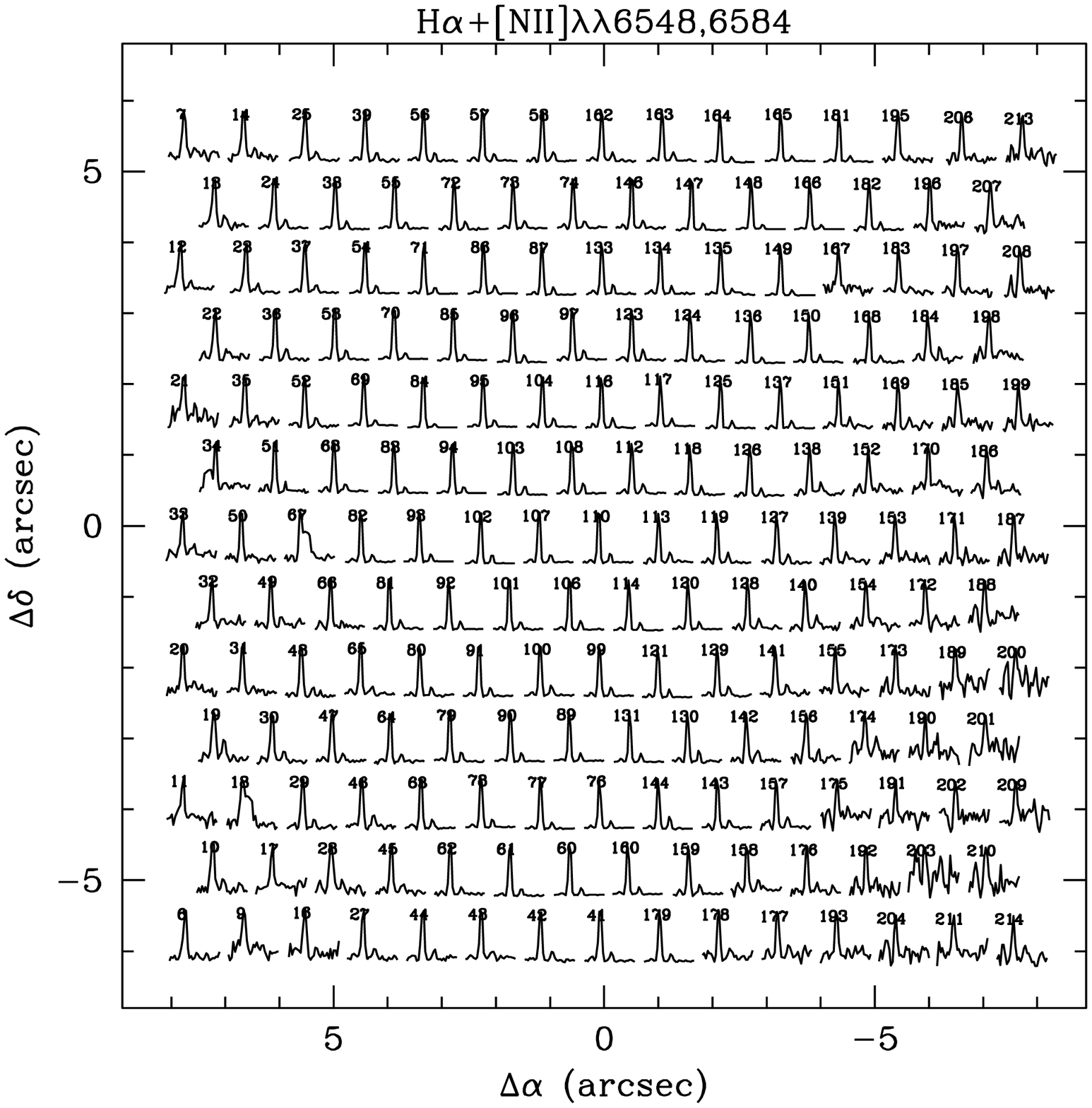,width=7.0truecm}}
\caption{Two-dimensional distribution of 
\Ha+[\ion{N}{2}]$\lambda\lambda6548,6584$ (6550--6640 \AA ).}
\label{apen4}
\end{figure}

\begin{figure}[H]                         
\centerline{\psfig{figure=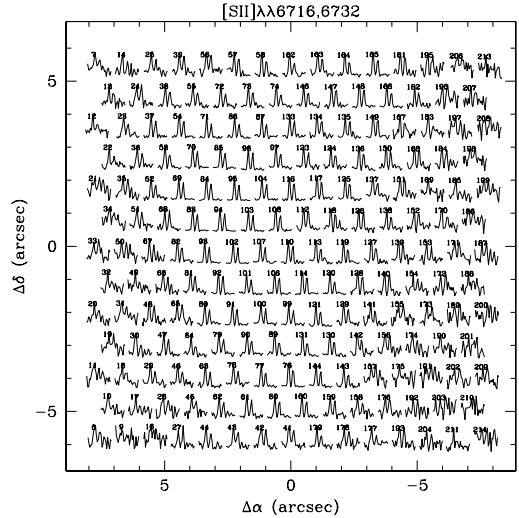,width=7.0truecm}}
\caption{Two-dimensional distribution of 
[\ion{S}{2}]$\lambda\lambda6716,6731$ (6720--6785) \AA ).}
\label{apen5}
\end{figure}

The nuclear spectrum of Mrk~370 derived from the INTEGRAL data is shown in 
figure~\ref{apen1}, in the whole wavelength range observed. 
We can easily recognize several emission lines. Here we only plot some of 
selected emission as spectrum diagrams 
(figures~\ref{apen2} through \ref{apen5}). 
Spectrum diagrams represent line profiles from individual spectra at each 
point in the observed region and in a small spectral range. The spectra at 
each location are normalized by the local peak so as to better display the 
profile shape (lines nearer to the bright knots are brighter than those 
farther out). 

\end{document}